# Statistical parameters of femtosecond laser pulse post-filament propagation on 65m air path with localized optical turbulence

Dmitry V. Apeksimov, Andrey V. Bulygin, Yury E. Geints, Andrey M. Kabanov, Aleksey V. Petrov, Elena E. Khoroshaeva

## Abstract

High-power femtosecond laser radiation propagates nonlinearly in air exhibiting pulse self-focusing and strong multiphoton medium ionization, which leads to the spatial fragmentation of laser pulse into highly-localized light channels – the filaments. The filaments are characterized by high optical intensity, reduced (even zero) angular spreading and can contain laser plasma or be plasmaless (postfilaments). The presence of optical turbulence on the propagation path dramatically changes pulse filamentation dynamics and in some cases causes pulse fragmentation enhancement and collapse arrest. For the first time to our knowledge, we experimentally and theoretically investigate the transverse profile of Ti:sapphire femtosecond laser radiation nonlinearly propagating a 65 m air path to the region of postfilament evolution after passing through an artificial localized air turbulence. We show that when a turbulent layer is placed before the filamentation region, the average number of high-intensive local fluence maxima ("hot points") in pulse profile as well as their sizes grow as the turbulence strength increases, and then saturates at some levels. On the contrary, the deposition of a turbulent screen within the filamentation region has almost no effect on both the number and the average diameter of the postfilaments.

## 1. Introduction

Propagation of high-power laser pulses of femtosecond duration through the gas and condensed medium occurs in a nonlinear mode and leads to significant transformations of spatial, angular and spectral characteristics of laser radiation [1]. The nonstationary self-action of ultrashort laser pulses in atmospheric air being inherently a process of nonstationary self-phase modulation of a light wave, is realized under the manifestation of many physical mechanisms, which can be in a local dynamic balance with each other. The key processes here are the beam self-focusing due to optical Kerr effect and the light-induced multiphoton ionization of the propagation medium. These processes are accompanied by nonlinear refraction and absorption of radiation by the self-induced laser plasma. This, in turn, clamps the further growth of pulse intensity and stabilizes the transverse dimensions of the high-intensity beam part leading to pulse self-channeling and the appearance of light filaments [1-3].

Thanks to the filamentation, a femtosecond laser radiation significantly enriches its spectral composition and becomes a source of a broad coherent radiation that can be promising for various atmospheric applications, in particular, for remote detection and physicochemical analysis of atmospheric aerosols [4]. Moreover, high power of femtosecond laser pulse promotes the realization of multiphoton processes in the medium such as the multiphoton ionization, stimulated fluorescence, super-radiance, which expands the capabilities of remote femtosecond diagnostics of atmospheric components also in the so-called, plasmaless postfilamentation regime of pulse propagation [5-9].

This specific pulse propagation regime, as recently shown experimentally [10-12], is realized after the termination of pulse filamentation and is not accompanied by active plasma generation in air. At the end of the filamentation region, a most intensive part of the laser beam, which formerly constituted the filaments, still preserves high energy concentration in the form of the narrow light beams (postfilaments) possessing high intensity (about several tenths of TW/cm$^2$) and reduced angular divergence in comparison with the beam as a whole. This is caused by joint action of Kerr self-focusing and specific annular pulse profile [13].

Importantly, similar high-intensive low-divergence light channels are also formed in air without any preceding pulse filamentation in the situations if the femtosecond pulse either has a subcritical or near critical power for self-focusing because of, e.g., energy partitioning into several sub-apertures [14], or initially possesses a complicated spatial amplitude-phase profile causing increased diffraction from the very start of propagation [15]. These localized light channels demonstrate similar quasi self-channeling as the postfilaments but differ from them in the spectral composition. Anyway, if such intensive light channels or postfilaments enter a denser (than air) medium during their free propagation, e.g., the particles of atmospheric aerosol or some solid objects, they can initiate recurring filamentation and plasma production in this denser medium [13, 15]. This may be of great practical interest for the femtosecond remote diagnostics and also opens up new prospects for using the postfilaments as a tool for the delivering of intensive broadband and highly directional radiation over the long ranges in the atmosphere.

Another important aspect in the problem of femtosecond radiation propagation along the long-range atmospheric links is the control over the spatial position and structure of the region where laser radiation maintains extremely high intensity, i.e., the region with the filaments and postfilaments. In this regard, the use of specially profiled radiation, i.e., laser beams with a transverse intensity distribution different from the Gaussian can be very promising. Due to the specific diffraction dynamics of such radiation during propagation in a nonlinear medium, the filamentation parameters also change and acquire an additional degree of freedom allowing one to taming the filamentation by changing the initial pulse profile. For example, tubular and combined so-called "dressed" beams [16] cause a noticeable filamentation delaying; a Bessel-Gaussian pulse profile leads to extending the filamentation region [14, 17]; the multimodal "coronal" type of energy distribution reduces the angular divergence of the postfilaments [18].

Practically, complex multimodal pulse energy distributions are produced by means of the amplitude-phase modulation of initial laser beam that results in optical splitting of the pulse energy into numerous spatially separated sub-beams of lower power [8, 14, 19, 20]. In this case, changing the number and configuration of the resulting beam maxima is usually achieved by means of various optical diffraction elements (phase plate, amplitude mask, diffraction grating, flexible mirror). At the same time, spatial modulation of the optical phase can also be obtained using the propagation medium itself by artificially creating areas of optical turbulence [21, 22].

Indeed, as a high-power femtosecond pulse propagates in real atmosphere in the filamentation regime, the unavoidable turbulent pulsations of air refractive index in the propagation path can significantly influence the filamentation dynamics. The onset of the filamentation region, its spatial extent and number of filaments become random [21-27]. On the contrary, the already formed filaments turn out to be robust against the stochastic distortions of laser pulse when passing through the turbulent region even when the turbulence significantly exceeds the atmospheric one [21, 27]. Note that in a certain situation atmospheric turbulence may also contribute to a higher spatial localization of the laser pulse than a still atmosphere [28].

Thus, by passing a femtosecond pulse with a regular phase through a turbulence layer it is possible to initiate stochastic fragmentation of the beam energy and its subsequent multiple filamentation in a certain section of the optical path [26]. Obviously, this will change the number of intense light channels formed after the filamentation region. So far, despite the large number of scientific papers on the subject of filamentation of ultrashort laser pulses in a turbulent atmosphere, the effect of optical turbulence on the statistics of the plasmaless postfilaments and methods to control their number remain unstudied.

This paper presents the results of our experimental studies, supported by the theoretical simulations of nonlinear propagation of phase-modulated high-power femtosecond laser pulses on an extended air path under the conditions of pulse Kerr self-focusing, plasma production and light filaments formation. The optical phase modulation is carried out by creating a spatially localized (about a meter thick) turbulent screen at different positions along the optical path using an industrial air heater (fan). The properties of the turbulent phase screen are controlled by changing the temperature of the air jet. Stochastic phase modulations of the laser pulse supported by the optical nonlinearity of the medium led to a small-scale self-focusing of the pulse, its spatial fragmentation and the appearance of many high-intensity light channels ("hot points") in the pulse profile at the end of the propagation. The subject of our research is to determine the dependence of the number and mean size of these light channels on the pulse energy, pulse shape and the spatial position of the turbulent screen.

## 2. Experimental methodic
### a. Schematics and measurement procedure

The experiments are carried out on an artificial air path with a maximal length of 120 m (only the first 65 m is used in the measurements) organized inside a building and containing no unaccounted sources of air turbulence. A Ti:sapphire laser with a carrier of 800 nm, pulse energy up to 35 mJ, duration of 105 fs, and 10 Hz repetition rate is used as an optical radiation source. A schematic of the experiments is shown in Fig. 1(a).

The experiments are carried out with two types of laser beams having a near-Gaussian and a multi-ring-shaped profiles. Initial beam diameter is 2.5 cm ($1/e^2$ fluence maximum) and different pulse energy $E$ controlled by the OPHIR-II power meter. The laser radiation is directed to a pivot mirror and is led to an airborne optical trace, where it experienced self-focusing and filamentation. Beam spatial structure is recorded with Pentax K-3 camera (equipped with a Pentax100MacroWR objective) and Andor Clara E CCD-camera on a white screen located 65 m away from the laser. The optical turbulence is formed as a heated air jet directed with a small angle (15-20 degrees) to the direction of femtosecond radiation propagation. To this end, the industrial turbo heater TESLA TH2200LCD is located at some distance downstream the pulse propagation. The temperature $T$ at the fan nozzle can vary stepwise from 100°C to 600°C with the maximum flow rate of 500 l/min.

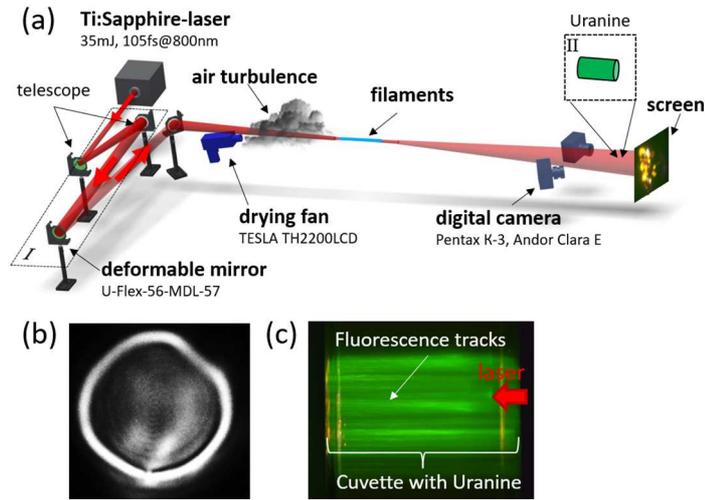

Fig. 1. (a) Experimental scheme for studying laser pulse filamentation modulated by a turbulent screen; (b) Pulse fluence profile after the deformable mirror; (c) Postfilaments emission in the uranine (side view).

For obtaining a multi-ring aberrational beam profile, an additional optical unit containing bimorph deformable mirror U-Flex-56-MDL-57 with 57 actuators arranged in several annular groups and a beam expander based on the Galilean telescope (unit group I in Fig. 1a) is placed in the optical circuit. By applying certain control voltages to the deformable mirror elements, a specific shape of the reflecting surface is created enabling the defocusing of beam central part and focusing of its peripheral area. As a result, after reflection from the deformable mirror and free propagation in air the laser beam acquires a multi-ring profile with gradual intensity decrease to the beam center as presented in Fig. 1(b). As shown in our previous works [13, 15], in contrast to an ordinary Gaussian beam such a ring pulse profile causes the regular multiple pulse filamentation along sufficiently long distance.

In the absence of air turbulence, the femtosecond laser pulse experiences the filamentation at the distances ranging from 54 m to 18 m when the energy in the pulse increases from 12 mJ to 35 mJ, respectively. The beam filamentation is undetectable in all cases when the thermal fan is turned on, except for the regime with $T = 100°C$ and $E = 35$ mJ when the onset of the filament approaches about 15 m. However, despite the absence of detectable laser plasma in the air optical path, multiple spatially localized regions with increased fluence always arise in the beam profile, which hereafter will be called the "hot points" (HPs). The fractional optical power in each HP is insufficient for the filamentation in air, but is quite capable of initiating the recurrent filamentation in a denser than air medium. Thus, in our studies we use a 10 cm glass cuvette filled with an aqueous fluorophore solution (Uranine WCC), which is placed at the end of the optical path immediately near beam profile registration area (group II in Fig. 1a). A picture of the uranine luminescence in the cuvette in the spectral region around 510 nm is shown in Fig. 1(c) when exposed to the laser pulse at the stage of postfilament propagation (after the end of the filamentation in air).

b. Optical turbulent layer characterization

Usually, in practice of atmospheric optical research, the theoretical conceptions of optical turbulence are used for the determination the turbulence strength in a medium by calculating the turbulence structural function in the framework of the "2/3" law in the Kolmogorov-Obukhov statistical theory. In particular, the structural function, or more specifically the structural constant of the refractive index $C_n^2$ can be measured from statistics of laser beam pointing walk-off when propagating through the turbulence by using the Kolmogorov or Karman spatial spectrum [21]. However, in the case of strong local artificial turbulence we used in a rather thin layer of heated air, this technique fails and leads to ambiguous results possibly because the turbulence produced by the fan heater used has a non-Kolmogorov spectrum [29].

In this regard, more successful method for $C_n^2$ characterization of the turbulent air jet turns out to be the measurement of the scintillation index (dispersion of intensity fluctuations $I$) $\sigma_I^2 = \left\langle \left( I - \langle I \rangle \right)^2 \right\rangle / \langle I \rangle^2$ of probe He-Ne laser (630 nm) after the propagation through a turbulent layer. A standard measurement scheme is used including a photodetector and a point diaphragm (0.1 mm diameter aperture) aimed at the center of the laser beam with 2 Hz data accumulation rate. The information collected over a selected time interval (2 minutes) is then statistically averaged. The examples of probe beam speckle structure are shown in Figs. 2(a) and (b) for two temperature regimes of the fan heater.

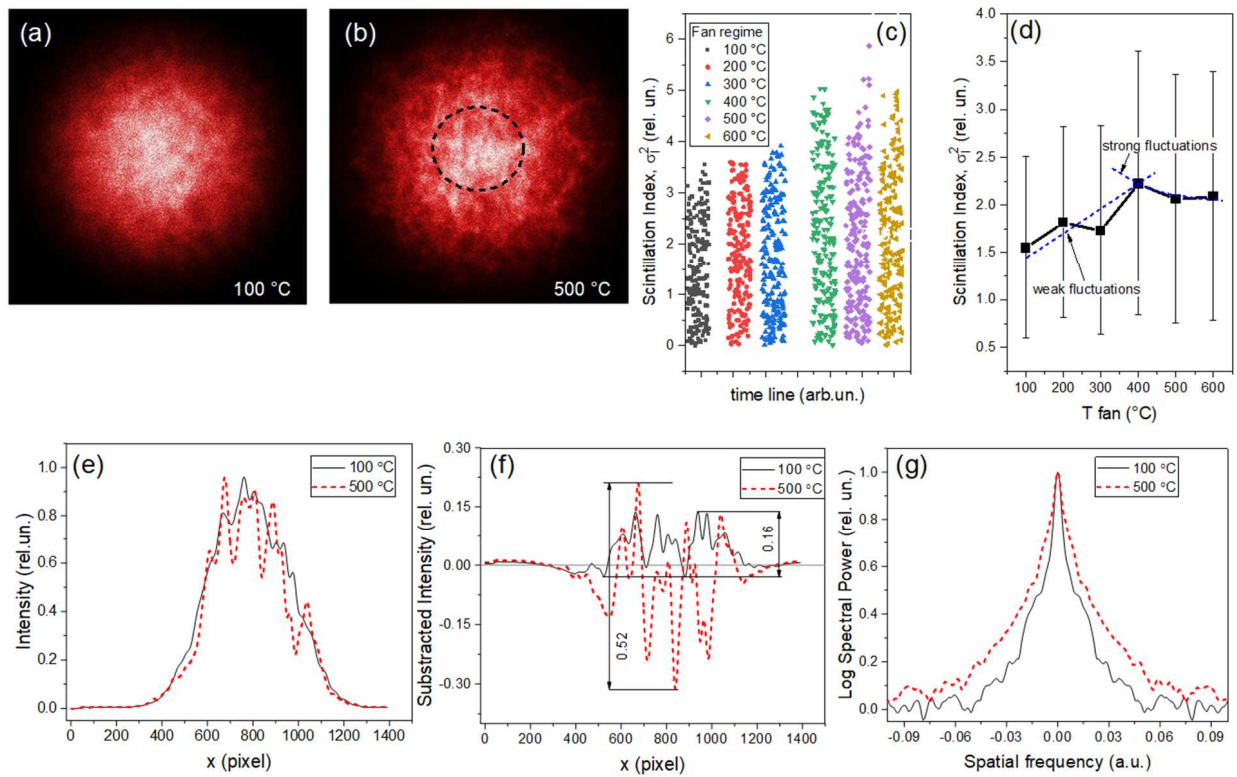

Fig. 2. Turbulence screen optical characterization. (a, b) Transverse profile of a probe He-Ne laser passed the turbulent layer at two fan temperature regimes $T$. (c) Scintillation index time series and (d) mean values versus fan temperature. (e, f) Intensity 1D-distributions at the beam center (shown by dashed circle in (b)), before (e) and after (f) initial Gaussian profile subtraction. (g) Beam power spectrum.

Fig. 2(c) illustrates the time series of the scintillation index $\sigma_I^2$ measurements in the central beam area, whereas Fig. 2(d) contains the data of statistical processing. As seen, the normalized turbulent intensity pulsations first grow up to the value $\sigma_I^2 \approx 2.25$ and then demonstrate a smooth decrease to $\sigma_I^2 = 2.05$ with the increasing of the conditional air jet temperature $T$. The extremum of the scintillation index is observed for $T = 400°C$. The character of the obtained dependence points to a certain analogy with the well-known result of Rytov theory for the behavior of the optical wave intensity dispersion in a random inhomogeneous medium [30]. According to this theory, two asymptotes can be distinguished in the dependence $\sigma_I^2(C_n^2)$ corresponding to the regimes of a weak, $\sigma_I^2 \propto C_n^2(T)$, and strong turbulence: $\sigma_I^2 \propto 1 + \left[C_n^2(T)\right]^{-4/5}$.

These asymptotes are plotted in Fig. 2(d) by dashed lines calculated through the least-squares fitting procedure. The calculations are performed according to the generalized phenomenological formulas taken from Ref. [31] for the scintillation index of a plane optical wave having a wave number $k = 10^7$ m$^{-1}$ propagating in a medium with the turbulence internal scale 5 mm (fan nozzle diameter). This allows one to estimate the conditional range of $C_n^2$-values, within which the turbulence strength of the fan air jet could change when switching the heater temperature regimes. It turned out that the thermal regimes switching from 100°C to 600°C conditionally corresponds to the change of structural turbulence constant in the range of $7 \cdot 10^{-9}$ m$^{-2/3} \leq C_n^2 \leq 3 \cdot 10^{-7}$ m$^{-2/3}$. As expected, these values are close to those given in previous similar studies on the femtosecond pulses filamentation [21, 27], but at the same time they are significantly higher than the typical values of atmospheric turbulence, where, as a rule, $C_n^2 \leq 10^{-10}$ m$^{-2/3}$. Strictly speaking, applying the classical theory of atmospheric turbulence for the characterization the optical parameters of a heated air jet is questionable. In this case, direct measurements of the structural function of refractive index fluctuations are required, but this is not the subject of our research. Moreover, since in our experiments the turbulence is concentrated in a meter-width air layer, one may conditionally consider some standard value atmospheric turbulence $C_n^2$ but distributed over the whole 65 m optical path.

A change in the fan temperature regime affects not only the temporal statistics of the laser beam intensity pulsations, but also its spatial spectrum. In Fig. 2(e) 1D intensity profiles (along the horizontal axis) are plotted after a test laser beam passes through a turbulent layer created at different fan temperatures. The intensity profiles shown in this figure are pre-processed using digital spatial filter for cutting-off the high-frequency noise components. It is clear that the amplitude of low-frequency perturbances increases sharply along with the fan temperature. This is more clearly illustrated by Fig. 2(f), in which the same intensity distributions are plotted but after subtracting the initial Gaussian beam profile. As seen, the amplitude of the low-frequency oscillations at fen temperature of 500°C is more than three times higher than its value for 100°C. This, in turn, affects the spatial beam spectrum (Fig. 2g) leading to its enrichment with increasing jet temperature, especially in the low-frequency region.

## 3. Results and discussion

Examples of the transverse energy distribution $F$ of femtosecond pulse at the distance $z = 65$ m from the laser source are shown in Figs. 3(a-e). Recall that the turbulent layer is created at a distance of 7 m even before the start of active plasma formation in air and pulse filamentation. All spatial profiles are digitally filtered and post-processed for better emphasize the fluence "hot points". To this end, the raw pulse images taken from the digital camera after passing the cuvette with uranine are first subjected to background noise subtraction at $1/e^2$ of the maximal value, then they are filtered using the Butterworth algorithm to cut-off high spatial frequencies and finally are normalized to the maximum. The fluorophore additionally attenuates weak HP and amplifies the most intensive "hot points", which are capable to produce filaments.

HP counting procedure is implemented automatically by the home-made FORTRAN code through the binarizing the resulting beam images and applying a standard connected area search algorithm. In Fig. 3, the number $N$ of HPs found is indicated at the bottom of each image. An increase of the air turbulence intensity occurring with the increase in air jet temperature, leads both to the increase in the number of intense fluence areas and to the transverse beam diffraction spreading in the region of the postfilament pulse propagation.

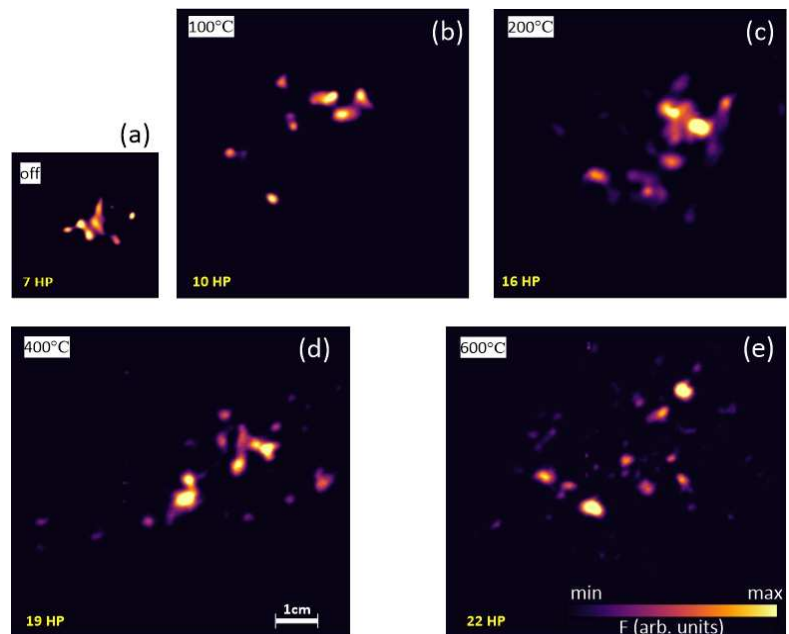

Fig. 3. Transverse pulse profiles (in false colors) with $E$ = 35 mJ at $z$ = 65 m (a) without and (b-e) after passing through the turbulent layer of different strength (top-left, indicated by fan $T$). The number of extracted HPs is shown with yellow numbers.

The results of the measured pulse profiles statistical processing at the end of a 65 m long optical path are shown in Figs. 4(a-c) as the dependence of HP number $N$ and diameter $D$ on the fan temperature regime $T$. Each experimental point is obtained by averaging approximately 20 independent measurements; the scatter of the corresponding value is shown by vertical lines.

From the analysis of these dependencies it follows that both the Gaussian and aberration (multi-ring) beam profiles exhibit a monotonous growth in the number of intense light maxima with increasing the strength turbulence of air jet (increasing $T$) and increasing laser pulse energy. At the same time, in the case of a Gaussian beam (Fig. 4a), the saturation of the HP number $N$ is clearly seen starting from the temperature of $T$ = 400°C, whereas in the multi-ring beam (Fig. 4b) this process occurs only for $T$ > 500°C. In addition, due to the initial multimodality of the intensity distribution, the very number of HPs in the aberration beam after the turbulent screen is on average almost twice as high as in the Gaussian beam. Worthwhile noting, in the situation without a turbulence, the mean number of HPs at the end of the optical path is always $N$ = 2 for a Gaussian beam and from 5 to 20 for the multi-ring profile depending on pulse energy. The addition of the turbulence raises these HP numbers to $N = 47$ in a Gaussian and from 29 to 86 in the ring beam.

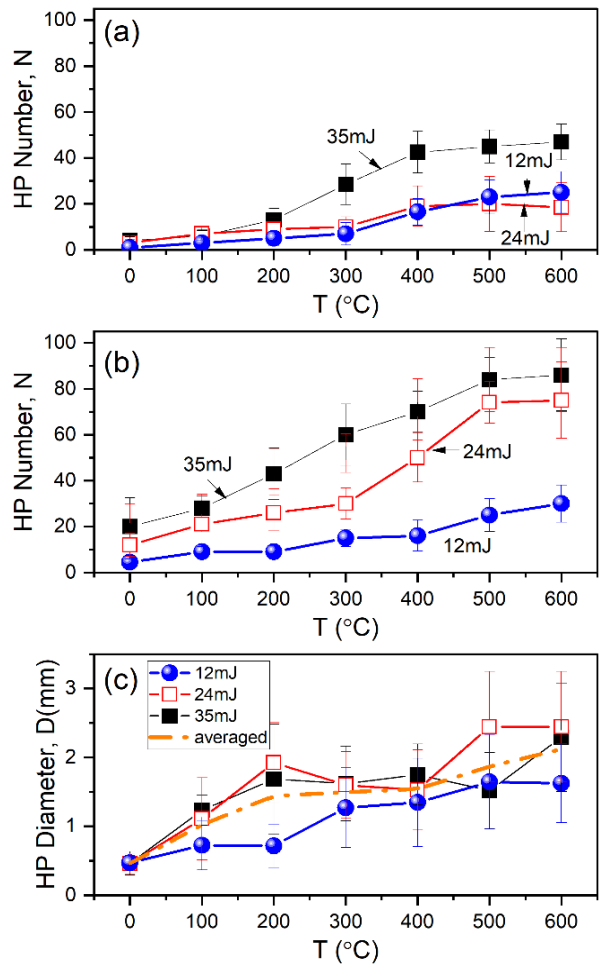

Fig. 4. (a, b) Number $N$ and (c) mean diameter $D$ of the «hot points» emerging in the laser pulse with (a) Gaussian and (b, c) multi-ring shape at different energy $E$ versus fan temperature $T$.

The measurements of the diameter $D$ of the bright points emerged in the laser pulse show that HP size varies within a fairly wide range and generally increases with the increasing of air turbulence strength, as shown in Fig. 4(c). Meanwhile, no unambiguous dependence of this parameter on laser pulse energy $E$ is revealed. On average, mean HP diameter increases approximately fourfold, from $D = 0.5$ mm in a still atmosphere to about 2.1 mm at the highest temperature regime of the fan. HP size increase is the consequence of the physical process of pulse diffraction enhancement after passing through a turbulent layer. Another cause is the technical one and relates the peculiarities of beam profile processing procedure, which does not allow distinguishing the neighboring HPs unless their mutual brightness contrast is low some threshold (merging of HPs). Important to recall, that despite a rather big size of the fluence speckles formed in the presence of a turbulent screen, the optical power carried by them can be sufficient to initiate the filamentation in a dense dielectric as the aqueous uranium solution (see, Fig. 1c).

Previously in [22] was shown, that for a high-power femtosecond pulse propagating in air with a localized optical turbulence, the position of the turbulent layer can have strong influence on the filamentation parameters. If the turbulent layer is placed after the start of pulse filamentation, it only weakly affects the onset and the total number of filaments. Otherwise, if the turbulent layer is placed before the start of filamentation, strong modulation instability of the pulse develops, which leads to significant position dispersion, instability, and delaying of the filaments up to the filamentation arrest [27].

We performed similar studies for the number of postfilament HPs formed in the laser beam at different spatial positions of the turbulent layer. The heater fan is placed at two positions: (a) before the filamentation area of the unperturbed pulse, which corresponds to the distance of 25 m, and (b) inside the filamentation region at the distance $z = 31$ m from the laser source. The results of these experiments for 24 mJ pulse are shown in Figs. 5(a, b).

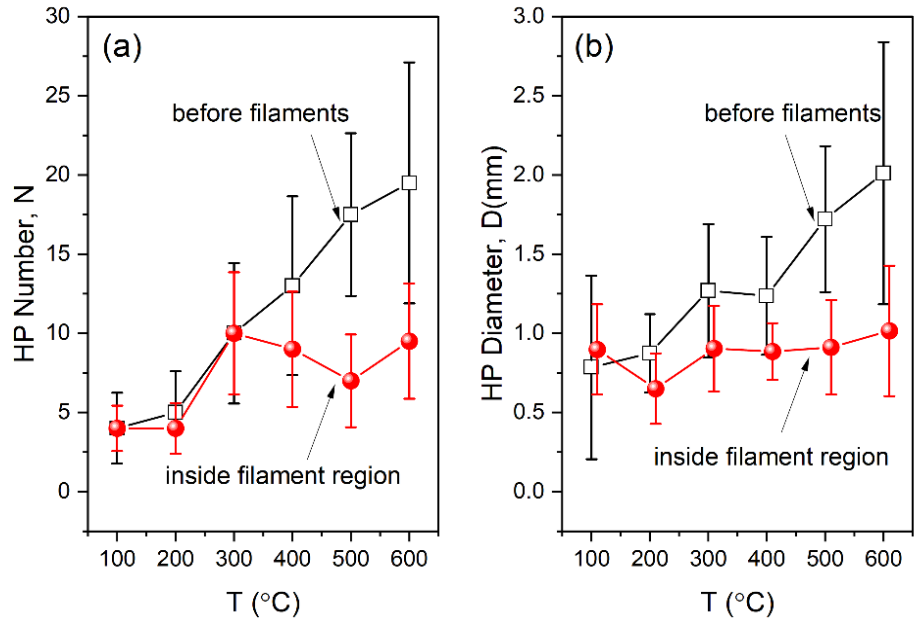

Fig. 5. Pulse postfilament statistics in the presence of local air turbulence. The dependence of (a) number $N$ and (b) mean diameter $D$ of the «hot points» on the fan temperature $T$ for laser pulse with $E = 24$ mJ (a Gaussian) and different turbulent layer position.

As is clearly seen, similarly to [22] by placing a turbulent screen in the region where the filaments are already formed has almost no effect on the number of postfilaments as the turbulence strength increases. This demonstrates the impressive stability of light filaments and postfilaments against the stochastic perturbations of pulse phase in the conditions of permanent maintaining a dynamic balance between the focusing (Kerr effect) and defocusing (plasma refraction) nonlinearities of the optical medium. Mean postfilament diameter shown in Fig. 5(b) is also almost constant in this situation and is about 0.8 mm at 50% variation. On the contrary, if the turbulent layer is placed prior to the pulse filamentation, an increase in both the number and size of the HPs is observed.

## 4. Numerical simulation of pulse filamentation in turbulent air by means of effective scatterers method

In this section, we discuss the results of our numerical simulations on nonlinear propagation of femtosecond laser pulses in the presence of a spatially confined layer with strong optical turbulence. Here, we consider the filamentation of laser radiation with the parameters close to the experimental ones, i.e., with a beam diameter of 7.8 mm and pulse energy up to 35 mJ. Numerical simulation of the filamentation of such sub-centimeter beams within the framework of the nonlinear Schrödinger equation (NLSE) or its analogs (FME, UPPE, etc.) [2], is challenging since it requires a fairly dense computational grid with a spatial step in the transverse direction smaller than 10 μm. This significantly complicates the implementation of large-scale numerical calculations [32].

Therefore, to reduce the computational resources with respect to laser filamentation we have developed an effective technique, in which the spatial region of the optical path where the filamentation takes place, is formally replaced by the region of pulse linear scattering on certain effective complex-valued phase screen. This effective phase screen (EPS) arises due to the joint action of optical Kerr effect and plasma refraction/absorption generated in medium during the pulse filamentation. This leads to the modulation of the effective air refraction index [1]. It turns out that the implementation of this effective approach to the filamentation simulation reduces the requirements for the spatial grid dimensions by at least an order of magnitude that allows one to perform parallel numerical calculations with the filament statistics accumulation even for centimeter-diameter beams propagating on real atmospheric paths.

Formally, the scattering of an electromagnetic field on EPS is presented in the following operator form:

$$U_{out}(\mathbf{r}) = e^{i\Psi(U_{in},\{\mathbf{R}\})} U_{in}(\mathbf{r}).\qquad(1)$$

Here, $U_{in}$ is the optical field before scattering on the screen, and $U_{out}$ represents the field scattered on the EPS $\Psi(U_{in}, \{\mathbf{R}\})$. The properties of this effective screen are given by the properties of the elementary light scatterers, i.e. plasma regions, which are a nonlinear functional of the incoming field $U_{in}$. In Eq. (1), $\{\mathbf{R}\}$ is a set of unknown parameters, which should be found on the basis of minimizing the residual function of the exact numerical solution via NLSE and the solution to Eq. (1) for a number of test pulse filamentation problems.

The particular analytic form of the test optical wave, i.e., the field Ansatz, is obtained from the well-known peculiarities of pulse single filamentation regime [32-35]. Thus, the Ansatz should describe the transverse structure of the output field $U_{out}$ as an annular field profile with high-intensity pedestal on beam axis and several low-intensity concentric rings, which sizes grow along with the pulse power.

Once a suitable Ansatz with a finite number of parameters $\{\mathbf{R}\}$ is selected, we one proceeds with the calculations of the residual vectors of resulting fields for a number of test examples by minimizing the difference of the output fields obtained on the sparse and dense numerical meshes based on certain norm. The search for optimal parameters of the Ansatz is a classical optimization problem belonging to the mathematical programming. Usually, this problem has many independent parameters, so commonly it is solved by a genetic algorithm (GA). For this purpose, we adopt a classical GA written on the parallel C-code [36].

To find the test solutions of the filamentation problem, the massive numerical calculations are performed in the framework of the NLSE formulated for the slowly-varying field envelope $U$ with a wave number $k_0$ at the carrier wavelength. The optical nonlinearity of air medium provides the pulse self-focusing and transverse collapse arrest by the plasma refraction. Thus, the slowly-varying envelope equation has the following form:

$$2ik_0 U_z(\mathbf{r}, z) = \left[ \Delta_\perp + \varepsilon_K (UU^*) + \varepsilon_m (UU^*)^{2m} + ik_0 \alpha_m (UU^*)^{2(m-1)} \right] U(\mathbf{r}, z) \quad (2)$$

Here, $U_z = \partial U / \partial z$, $\Delta_\perp$ stays for the transverse Laplace operator, $\varepsilon_K$ is the nonlinear additive to the medium permittivity accounting for the Kerr effect, which causes optical wave self-focusing. $\varepsilon_m$ represents the additive due to action of $m$-order optical nonlinearities which stop the pulse collapse by saturating the focusing term $\varepsilon_K$. This additive accounts for the wave refraction of plasma areas. Lastly, the term $\alpha_m$ accounts for pulse energy losses due to $m$-photon ionization of air molecules and plasma filament absorption.

As the initial optical field, we chose a set of $N_g$ Gaussian pulses $U_n$ with the amplitudes $A_n$ and radii $r_n$ ($r_n \leq 1$ mm):

$$U_n(r) = A_n \exp\{-(r/r_n)^2\}, \text{ at } n = 1 \ldots N_g \quad (3)$$

The partial power of the pulses, $P_n \sim A_n^2$, are uniformly distributed in the range $1 \leq P_n/P_c \leq 10$, where $P_c$ is the critical self-focusing power in air (~ 4 GW). These specific ranges of the pulse power and radius are chosen due to the typical sizes of field perturbations [37] giving rise to the single filament when wide-aperture laser beams experiences multiple filamentation in the atmosphere.

We introduce a set of spatial field moments, $Q^{nl} = \int (U_n U_n^*)^{2l} |\mathbf{r}|^{2l} d\mathbf{r}$, which is used for carrying out the optimization sweep and minimization of the residuals between different solutions. If the criteria $Q^{ml}$ are minimal, then the numerical solutions of the problem (1) obtained using the EPS method matches in the best way the test solution of the NLSE (2). Usually, the required number of field moments is about ten. Separate solutions are realized at a distance of 30 cm upon reaching a certain preset sub-filamentation level of pulse intensity (~ $10^{12}$ W/cm$^2$), where the effective screen is activated.

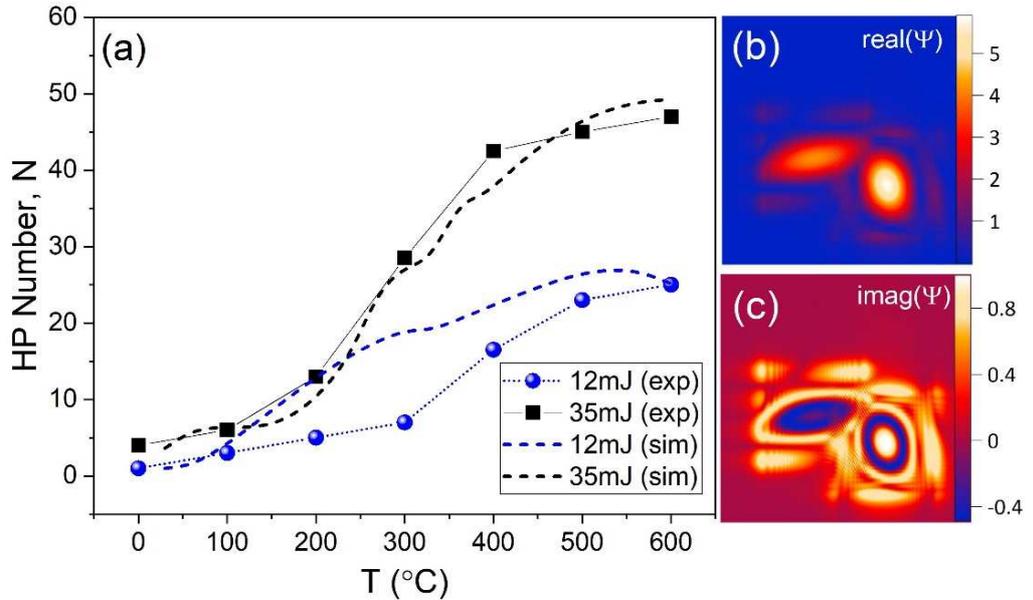

Fig. 6. (a) Comparison of experimental measured (*exp*) and numerically calculated (*sim*) HP number $N$ emerged in a femtosecond pulse with different energy at different fan temperature $T$. (b, c) The real (b) and imaginary (c) effective phase screens (EPS) used in the simulations of Gaussian pulse multiple filamentation.

Upon the implementing the EPS methodic, the initial genome $\Psi(U_{in}, \{\mathbf{R}\})$ generation is performed randomly within the data set $\{\mathbf{R}\}$ including available in the literature pulse filamentation experimental data. GA used in our simulations has the initial population depth of 40 "species" with uniform crossover and roulette wheel selection of the progeny. As a result of the GA, the most suitable EPSs are determined to mimic the air filamentation patterns of a laser pulses with a 1cm radius and an energy from 10 to 35 mJ. These complex-valued phase screens stop the field amplitude growth in the self-focusing regions of beam and, additionally, provide the required (adequate to the full model) physical mechanism for diffraction spreading of both the postfilaments and their vicinity. As an example, the real and imaginary parts of the EPS constructed for the experimental laser pulse with the energy 35 mJ is shown in Figs. 6(b,c). Note two main areas of maximal complex refractive index perturbations simulating two separate filaments observed with the experimental pulse of a Gaussian shape.

According to the in-situ experiments, the presence of a turbulent air layer on the optical path should be also reproduced. To this end, we modify the equation (2) by including a random permittivity screen $\varepsilon_{tb}$ along with the effective complex screen $\Psi$, as follows:

$$2ik_0 U_z(\mathbf{r}, z) = \left[\Delta_\perp + \varepsilon_K(UU^*) + \Psi(U_n(0), \{R\}) + \varepsilon_{tb}\right] U(\mathbf{r}, z) \quad (4)$$

The turbulent phase screen is constructed by the spectral method described elsewhere [23]. To simplify and speed-up the numerical calculations, the spectral density function of turbulent pulsations $\varepsilon_{tb}(\mathbf{k})$ is modeled by a step function: $\varepsilon_{tb}(\mathbf{k}) = \varphi_0 \theta(k - k_{high})$, where $k_{high}$ is the upper limit of the spatial frequencies of the turbulence, and $\theta(k)$ is the Heaviside function. The spectral amplitude of the turbulence $\varphi_0$ is conditionally associated with the temperature of the heating fan $T$ according to the relation, $T = \nu \varphi_0 k_{high}^2$, where $\nu$ is certain free parameter which is determined from the condition of best correspondence between theory and experiment. A comparison between the experimental and obtained in our simulations dependencies of the "hot points" number $N$ formed in a Gaussian beam of different energy at the postfilament propagation in air is presented in Fig. 6(a).

As seen, generally there is a good agreement between the experiment and the theory which gives the correct quantitative behavior of the parameter $N$ with the variation of air turbulence strength and pulse energy. From this figure, two basic relationships are evident. First one is the growth of the of HP number along with the pulse energy (power) increase that is described by the known functional trend relating the number of filaments and pulse power by a direct proportion [2].

Second relationship demonstrates an increase in the "hot points" number with increasing heater fan temperature followed by the smooth saturation of $N$. This relationship is realized in a certain range of scales of field inhomogeneities and pulse power, and as follows from Fig. 5(a), also depends on the position of the turbulent screen. Qualitatively, this can be explained by the Bespalov-Talanov modulation instability theory [37], which shows that the perturbations of an optical wave phase leads to HP number increase in wave amplitude profile when propagating in a Kerr medium. However, only those HPs will be supported by self-focusing optical nonlinearity, in which the nonlinearity suppresses the diffraction, i.e., which carries enough optical power. In our theoretical model, an increase in the fan temperature causes a stronger fragmentation of pulse profile. Consequently, at a fixed energy in the pulse starting from some values of $T$, an increase in the air temperature will lead only to the appearance of weak low-intensity ripples in the beam fluence profile. However, these energy perturbations do not contain the necessary critical power to initiate pulse collapsing due to the self-focusing, and hence a saturation of the HP number is observed.

## 5. Conclusion

To conclude, we present the results of the experimental and theoretical studies on the multiple filamentation of high-power femtosecond pulses of a Ti:sapphire laser (800 nm) in extended air path (65 m). To induce multiple filaments and postfilament light channels during the pulse nonlinear propagation, we applied strong pulse phase modulation by depositing an artificially created turbulence screen in the form of a spatially localized (about 1 m long) jet of heated air. This turbulent layer is created with an industrial heating fan and placed at different points in the optical path both before and inside the pulse filamentation region. After passing the turbulent layer and self-focusing in air, the laser radiation acquires a transverse profile consisting of many bright randomly distributed light spots ("hot points") representing local maxima of pulse energy density. The energy in every such maximum is still sufficient to initiate filamentation in the aqueous solution of the fluorophore (uranine).

We studied the statistics of these "hot points" at the end of the optical path on the stage of pulse postfilament propagation. The dependencies of the number and size of the "hot points" on pulse energy, initial spatial profile (Gaussian, multi-ring), the position and strength of the turbulence layer are obtained. We found that the increase of the turbulence in air layer leads to the increase in the number and average diameter of "hot points" in the beam profile if the turbulence layer is located before the pulse filamentation onset. This increase is most pronounced for the multi-ring beam due to the initial pulse energy partitioning into several rings. At very strong turbulence, the "hot points" number tends to saturate. Positioning the turbulent screen inside the pulse filamentation region practically does not change the statistical parameters of the transverse beam profile.

The theoretical simulations of a femtosecond pulse filamentation in turbulent air are carried out using the originally-developed method of effective light scatterers, which formally replaces the filamentation region by a complex phase screen, whose structure is determined using machine learning techniques (a genetic algorithm). A good agreement is achieved between the experiment and the theory, which gives a correct trend in the number of formed "hot point" with the changes in the heating fan temperature and laser pulse energy.

We propose the findings presented can be useful for the problems of remote femtosecond diagnostics of atmospheric constituents (liquid aerosol, dust) by implementing pulse fragmentation into many low-energy parts before the filamentation using a thin but strong turbulent phase screen. This can destroy or even prevent the pulse filamentation and produce instead many plasmaless postfilaments possessing yet high intensity sufficient for inducing the multiphoton processes (ionization, fluorescence) in, e.g., aerosol particles.

## 6. Funding